# Output stages inside a negative feedback loop: application to a low-voltage three-phase DC-AC converter for educational purposes

FRANCISCO LLOPIS [(a)] and MARIO JAKAS [(b)]
[(a)] Departamento de Física Fundamental y Experimental, Electrónica y Sistemas
Universidad de La Laguna
38205 La Laguna, Tenerife
SPAIN
[(a)] fllopis@ull.es   [(b)] mmateo@ull.es

*Abstract:* - The circuit presented in this paper aims at providing three 40 $V_{pp}$ 50Hz AC voltages sources with 120º phase separation between them. This is a fully analogue circuit that uses standard, low-cost electronic components without resorting to a microcontroller as previously proposed by Shirvasar et al [1]. This circuit may serve as a basis for a low-voltage 3P-AC power supply that students may safely use to realize experiments, i.e. about the principles and applications of three-phase AC power lines, without the risk of electric shocks.

## 1 Introduction

The lack of equipments which can be safely used in demonstrating the operation of three-phase systems without the risk of electric hazards poses a serious limitation to first courses of electrical and electronics engineering and science degrees. Consequently, instructors are normally forced to teach these subjects by resorting to theoretical demonstrations, animated graphs and, sometimes, computer simulations. But the absence of laboratory experiments leads students to have a rather poor understanding of electrical and electromechanical devices behaviour. At best, these types of laboratory experiments are carried out under a strict supervision and a carefully controlled situation, so that students do not actually conduct the experiment by themselves.

In an attempt to overcome this problem, various electronic circuits have been proposed [1-4]. Among them, it should be mentioning that of Shirvasar et al. [1] who have recently proposed a microcontroller-based circuit capable of generating three signals with a 120 degree phase-separation. The voltage signal is obtained using PWM technique and the output stage was implemented with six power MOSFET transistors mounted on an H-bridge configuration. The PWM pulses acting on the MOSFET gates are properly furnished by a 16F686 microcontroller.

After assessing the usefulness of such an idea and considering that similar low-cost commercial equipment was not available, we built the circuit described in Ref.[1] and several difficulties appeared. In the first place, the signals were slightly distorted, perhaps, due to the poor sampling and gate-driving signal of the MOSFET. Secondly, the quality of the output AC voltage appeared to depend on the load. Needless to say that the circuit so mounted was able to have a three-phase motor, previously constructed by us, properly running. But the task of improving the quality of the output voltage remained unsolved.

It turned out then, that one can conceivably replace the microprocessor-driven PWM system for an analogue oscillator and, similarly, substitute the H-bridge for three, conventional power amplifier. In fact, a three-stage 120 degree phase-shift oscillator seems to be suitable for the present circuit and, as output stages, three BJT or MOSFET push-pull followers preceded by three linear preamplifiers may work fairly well.

## 2 Problem Formulation

Fig.1 shows the block diagram of the proposed analogue circuit. It comprises an op amp based phase-shift oscillator capable to generate three 250 $mV_p$ and 50 Hz sinusoidal signals with a phase separation of 120º. As depicted in the same figure the circuit is feeding three loads connected in the delta configuration. Although not shown in Fig.1, loads can also be connected in the star configuration. But the system is also intended to develop three amplitude stabilized signals $v_A$, $v_B$ and $v_C$ even when connecting low-impedance loads, and thus the need of employing three power stages arises.

## 2.1 The basic configuration

Bearing in mind that -at least theoretically- negative feedback helps to stabilize amplifiers gain [5], we explored first the circuit shown in Figure 2. This configuration, a non-inverting voltage amplifier connected to a power stage, both inside a negative feedback loop, is commonly introduced in several textbooks that cover the basic analogue building blocks [6, 7]. Although a power op amp can accomplish the same task, in our first attempts we decided to build a more affordable circuit employing low-power op amps and power BJTs.

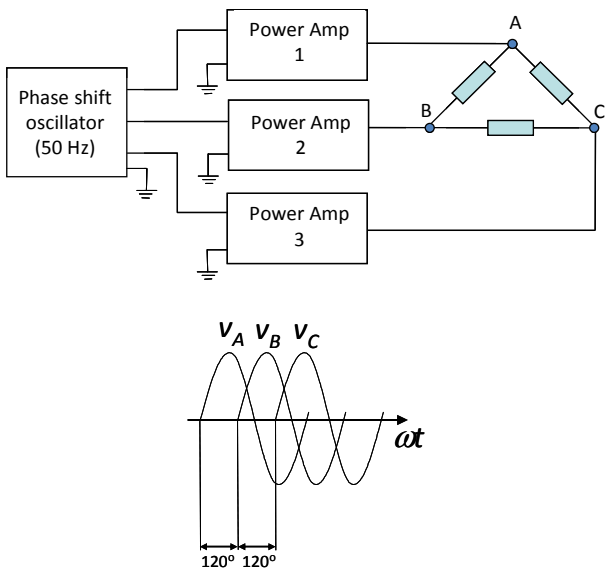

Fig. 1 Block diagram of the DC/3P-AC converter circuit. A three-phase oscillator delivers three, 120º phase-shifted sinusoidal signals, to three independent power stages.

First, it is usually shown that class B amplifiers - like those built in the push-pull configuration- exhibit a greater efficiency than class A ones: thus, the former are often preferred as output stages in audio amplifiers, for example. Besides, transistors operate as emitter (or source) followers boosting the current provided by the op amp.

However, the output signals of class B output stages are affected by the so-called crossover distortion, an intrinsic effect in transistors when driving them from cutoff to active operation mode. But a nearly pure sinusoidal signal can be obtained by introducing negative feedback around the open loop amplifier, overcoming the effect of crossover distortion. The open loop amplifier consists in an op amp driving the push-pull output stage, which does not introduce phase shift since transistors are connected as emitter (or source) followers. Therefore, there is no phase shift between the input voltages of the op amp, as expected for amplifiers with negative feedback. In this fashion, provided that an ideal op amp has an infinite differential gain, it is fulfilled the condition of a virtual short-circuit between the op amp inputs. This fact, together with the assumption of negligible op amp input currents, ensures that the overall gain is mainly determined by the feedback resistors as $1+R_f/R_i$, a result that is highlighted in the aforementioned textbooks.

If $v_i$ represents one of the sinusoidal signal generated by the oscillator, the voltage output developed across the resistive or reactive load should be $v_o = (1 + R_f/R_i) v_i$, which is a pure sinusoidal waveform too.

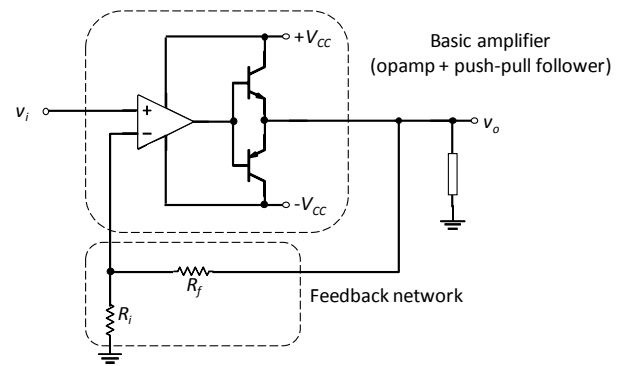

Fig. 2 Basic configuration: a non-inverting voltage amplifier with current-boosting capability has an overall voltage gain determined by the feedback resistors.

## 2.2 Operation with resistive loads

In our first attempt, the circuit operated with voltage supplies of ±18V. Each power amp was built employing an LF411 IC and a pair of complementary transistors BD437/BD438. Three resistances rated at 100Ω/5W were connected to each power stage in the star configuration. For 250 mV$_p$ oscillator voltage outputs, $R_i$ and $R_f$ were picked to obtain about 12 V$_p$ at the output of each power stage. The three signals exhibited the same amplitude, as can be expected when loads are balanced.

## 2.3 Operation with RL loads

The second attempt consisted in feeding a three-phase motor as indicated in figure 3. Each phase exhibited an equivalent impedance of (36 + $j$180) Ω at 50Hz. The same voltage levels were obtained in this case and the circuit was capable to maintain the motor running. But after turning off the power supplies, we observed that a pair of complementary BJTs got burnt. We attributed this effect to the magnetization of the coils, leading the operating point of the BJTs outside the safe operating area (SOA).

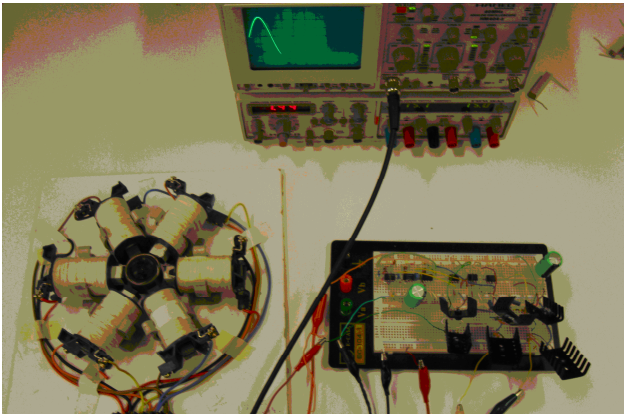

Fig. 3 Picture of the three-phase DC-AC circuit supplying a 3P-inductor motor used in classroom demonstrations.

### 2.3 Increasing the output signals amplitude

Another drawback of the circuit shown before relies on the maximum voltage swing of the op amp. We attempted to increase the voltage amplitude of the op amp driver by employing a suspended supply scheme as proposed in reference [7]. Although not shown, the circuit worked fairly well with BJT output stages and resistive loads, but the failure with RL loads remained.

## 3 Problem Solution

In a third attempt we tried to drive each RL load with a class B power MOSFET stage driven by a LM675 power op amp (figure 4). This device can be operated with supply voltages up to ±30V. Although low-power high-voltage op amps seem more suitable for this purpose, we did not employ them because they are more expensive. Since we had some units of the LM675 power op amp available, we decided to employ them to drive each MOSFET push-pull stage (fig. 4) The circuit worked properly in successive trials and MOSFETs were not damaged after turning off power supplies. We also noticed that, when connecting capacitive loads, the currents drawn from the supplies didn't reach stable values. In order to correct this problem each circuit finally included small source resistors to avoid thermal drift.

## 4 Conclusion

A fully analog, low-cost, three-phase generator is built. It is based in an op amp phase-shift oscillator plus three power stages. The latter are three push-pull stages inside a feedback loop which determines the voltage gain. This circuit was proven to work remarkably well for classroom demonstrations of three-phase electric power and also illustrate the benefits introduced by negative feedback.

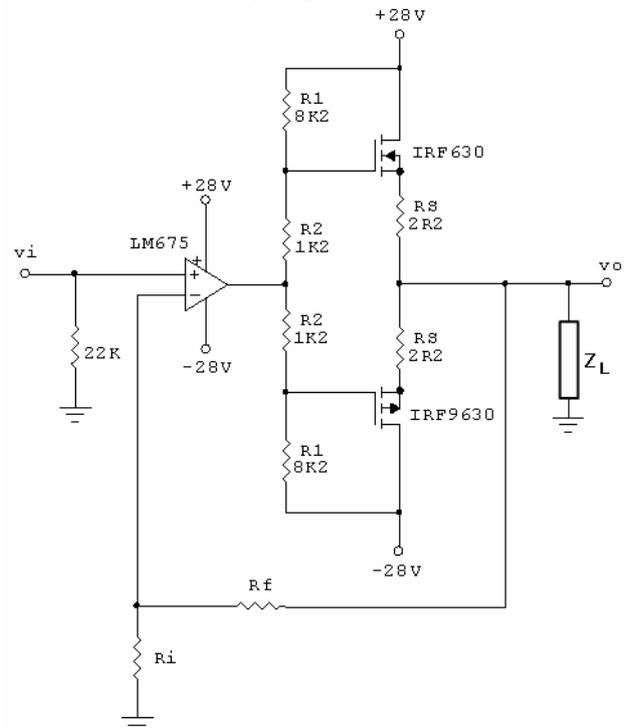

Fig. 4 Power amplifier schematic: MOSFETs are used in the push-pull stage, which is driven by a LM675 op amp.